\documentstyle[sprocl,epsfig]{article}

\def\be{\begin{equation}}
\def\ee{\end{equation}}
\def\bea{\begin{eqnarray}}
\def\eea{\end{eqnarray}}

\begin{document}

\rightline{\vbox{\halign{&#\hfil\cr
&SLAC-PUB-10853\cr
&IPPP/04/72\cr
&DCPT/04/144\cr}}}
\vspace{1cm}

\title{The Linear Collider Physics Case:
International Response to the Technology Independent Questions
Posed by the International Technology Recommendation Panel\\}

\author{ Klaus Desch$^a$, JoAnne Hewett$^b$, Akiya Miyamoto$^c$,
Yasuhiro Okada$^c$,\\
 Mark Oreglia$^d$, Georg Weiglein$^e$, 
Satoru Yamashita$^f$ }

\address{$^a$Universit\" at Hamburg, Institut f\" ur Experimentalphysik, 
Luruper Chaussee, D-22761, Hamburg, Germany\\
$^b$Stanford Linear Accelerator Center, 2575 Sand Hill Rd,\\ Menlo Park, CA
94025, USA\\
$^c$High Energy Accelerator Research Organization (KEK), Tsukuba, Ibaraki 
305-0801, Japan\\
$^d$Enrico Fermi Institute and Department of Physics, University of
Chicago, Chicago, IL  60637, USA\\
$^e$Institue for Particle Physics Phenomenology, University of Durham,\\
Durham DH1 3LE, UK\\
$^f$International Centre for Elementary Particle Physics and Department
of Physics, University of Tokyo, Tokyo 113-0033, Japan\\
}


\maketitle\abstracts{The International Technology Recommendation Panel
distributed a list of questions to each major laboratory.  Question
30, part b and d, were technology independent and  related
to the physics goals of the Linear Collider.
An international panel, with representation from Asia, Europe, and
the Americas, was formed by the World Wide Study during LCWS04 to formulate
a response.  This is given below and constitutes the response
of the world-wide Linear Collider community.
}

\noindent{\bf {30b)
How do you make the case for determining the final energy 
choice for the LC prior to the LHC results?  What if LHC results 
indicate that a higher energy than design is required?}}
\medskip
 
The physics case for the $200-500$ GeV Linear Collider, upgradable to 
energies around 1 TeV, rests on arguments that are independent of the 
findings at the LHC. (We note that this design and upgrade energy are 
common to both the warm and superconducting technologies.)  There are 
many reports that document this physics case.  We cannot repeat all of 
these documented arguments, so will only recall the essential points 
here.  The question of whether the top end energy should be 800 or 1000 
GeV will be commented on at the end of this response.

1) Electroweak Symmetry Breaking:  The LC will decipher the mechanism 
responsible for electroweak symmetry breaking, regardless of whatever 
it may be.  If the Standard Model is a good low energy effective theory, 
the current precision electroweak data indicate that the Higgs boson is 
lighter than about 250 GeV.  In addition, supersymmetric extensions to 
the SM also predict that a light Higgs boson exists.  If the Higgs is 
similar in nature to that predicted by the Standard Model, it will be 
discovered at the LHC and a LC will be essential in order to study its 
properties in detail in a model independent way.  The precision 
measurements of the Higgs couplings available at the LC  will distinguish 
the Standard Model Higgs from those which can arise in many other 
scenarios.  In particular, the 500 GeV LC can precisely measure the 
elementary couplings of the Higgs boson to quarks, leptons, and gauge 
bosons, and the spin and CP properties of the 
Higgs boson may be determined via a 
threshold scan.  These measurements are essential for an experimental 
verification of the scalar dynamics underlying the electroweak symmetry 
breaking.

Some scenarios predict that a light Higgs may decay in such a way
that it escapes detection at the LHC.
By utilizing the recoil mass technique
in the reaction $e^+e^- \to Z+$Higgs will allow for a Higgs discovery at the 
LC in a model independent fashion.  

If the Higgs boson is heavier than indicated by the precision 
electroweak data, its properties can be accurately determined at a 1 TeV 
LC.  Furthermore, consistency with the precision electroweak data 
implies that other new particles whose masses lie below 1 TeV must also 
be present.  As discussed below, the LC then plays a crucial role in 
identifying the nature of this new physics.

A last possibility is that a light Higgs boson is not realized in nature. 
In this case, $WW$ scattering violates unitarity at around a TeV unless 
there is new physics. If the associated new states would be out of direct 
reach, precision measurements would enable the LC to observe the effects 
of these states through their virtual contributions to existing processes. 
This has been studied in great detail for many scenarios, such as strong 
electroweak symmetry breaking.  In particular, the complete threshold 
region of the new strong interaction can be explored at a LC with energies 
around 1 TeV.

2) The Hierarchy Problem:  A shortcoming of the Standard Model is its 
instability against the huge hierarchy of the vastly different scales 
relevant in fundamental physics.  The Higgs and gauge boson masses are 
unstable to quantum fluctuations and would naturally rise to the Planck 
scale without the onset of new physics around a TeV. It is essential 
that this hypothesis be tested as precisely as possible. The leading 
candidates for resolution of this hierarchy problem are:
  
  (i) Weak-scale Supersymmetry.  The lightest superpartners are expected 
to be within reach of a $500-1000$ GeV LC. A LC with its capabilities of 
polarized beams and threshold scans will precisely determine the properties 
of this spectrum.  Only the combined information of the LHC and LC 
measurements can decipher the supersymmetry breaking mechanism and provide 
clues about the physics at the grand unified scale.
 
  (ii) Extra Spatial Dimensions.  In this case, a 1 TeV LC can observe 
both the direct production and virtual effects of Kaluza-Klein 
excitations of the graviton and Standard Model particles. The LC with its 
superb energy resolution does a particularly good job observing the 
narrow resonances characteristic of some models. Including virtual effects, 
the discovery reach of a 1 TeV LC for these excitations is $6-20$ TeV and 
will cover the natural region of parameter space that is relevant for 
resolution of the hierarchy problem. 
 
  (iii) Little Higgs Models.  These models predict a strongly interacting 
sector and predict the existence of new scalars, gauge bosons, and fermions 
at energies of order 10 TeV.  The sensitivity to physics that lies beyond 
the direct energy reach of the LC can provide important clues to the high 
energy behavior.  In particular for this case, the LC can determine the 
couplings of these new particles to the Higgs sector and verify the specific 
structures of such models. 

In any scenario addressing the hierarchy, precision measurements of Standard 
Model processes at the LC with polarized beams are sensitive to virtual 
effects at high energy scales and will be crucial to determine the nature 
of the new physics.

3) Dark Matter:  One of the simplest explanations for cosmic dark matter, 
the invisible matter that constitutes 80\% of the mass of large clusters 
of galaxies, is that it is composed of a new stable particle with weak 
interaction cross sections.  Astrophysical observations are consistent with 
the mass for such a particle being of the order of 100 GeV and it would thus 
be copiously produced at the LC.  In this case the LC would be ideally 
suited for establishing the quantum numbers of dark matter candidates; this 
is discussed more in the answer to 30d. 

4) Precision measurements of the Standard Model:  A 500 GeV LC will make 
important precision measurements within the Standard Model. (i) The mass 
of the top quark can be measured to an accuracy better than 100 MeV and 
the top quark couplings to the photon and the $Z$ can be determined at the 
percent level.  The uncertainty on the top quark mass is a limiting factor 
in the global fit to the electroweak data set.  A measurement with 100 MeV 
precision, together with the improved measurement of the $W$ mass at the LHC 
(or with even better accuracy by running the LC at the $W$ pair threshold), 
thus allows for much better exploitation of the LHC results and a precise 
consistency check of the Standard Model with unique sensitivity to new 
physics beyond 1 TeV.  (ii)  The LC can measure the $WW\gamma$ and $WWZ$ 
couplings to parts in $10^4$.  The radiative corrections to these couplings 
within the Standard Model are at the level of $10^{-3}$.  The LC thus has the 
sensitivity to probe new physics contributions at a high level of precision.  
In particular, these are key experiments that are sensitive to new strong 
interactions in the Higgs sector.  (iii)  The LC can determine $\alpha_s$ to 
better than 1\%, and a precise evolution of 
$\alpha_s$ is an important ingredient 
for models of grand unification.  (iv)  The option exists to run the LC at 
the $Z$-boson pole, and at the $W$-boson pair production threshold.  The high 
luminosity of the LC will allow for $10^9$ $Z$-bosons to be produced.  This 
Giga-$Z$ option will allow for the measurement of the effective weak mixing 
angle at the $10^{-5}$ level (an order of magnitude 
improvement), of the $W$ boson 
mass to $6-7$ MeV, and of $\alpha_s$ 
to 0.4\%.  Together with the 100 MeV top mass 
determination, this will be an unprecedented precision test of the Standard 
Model, which would be all the more important in the unlikely event that the 
LHC discovers nothing.  

The arguments stated above demonstrate the need for a $500-1000$ GeV LC, 
regardless of LHC results.  If the LHC experiments only discover a particle 
sector at mass scales beyond 1 TeV, it will be important to establish the 
effects on Standard Model processes via precision measurements, and to search 
for lower mass states which might have couplings or backgrounds which would 
prevent their observation by LHC experiments.  In many cases, the sensitivity 
to new physics via virtual effects at the LC exceeds that of direct searches 
at the LHC. Precision EW measurements by a LC are important for distinguishing
among multiple interpretations of new physics which may be observed by the LHC.

It is difficult to make a strong case for whether the top energy of the LC 
should be 800 GeV or 1 TeV.  Certainly, the higher energy provides a somewhat 
higher window to new physics and gives larger production rates for some 
Standard Model processes, such as those relevant for the Higgs self-coupling 
determination.  However, there is an energy-luminosity tradeoff which also 
must be considered for the different processes.

The combined knowledge gained from both the LHC and LC programs will be 
necessary to make qualified decisions about post LHC/LC facilities.  In
particular, the experience of operating of a $0.5-1.0$ TeV LC will be crucial
to outline the design of a multi-TeV machine.
\bigskip

\noindent{\bf {30d)
Considering the LC will start much later than LHC (although it can have a 
concurrent operation period), what physics capability does LC have which LHC 
does not share?  Can this be realized at 500 GeV or does it require much 
higher energy? }}
\medskip

The LHC and the LC have complementary and synergetic physics capabilities. 
This synergy can be best explored if both machines run concurrently.  However,
the LC has unique physics capabilities that are crucial to our understanding 
of nature and will be needed regardless of the LHC findings and the LC 
startup time. The LHC strength lies in its mass reach, while the LC is a 
precision machine with:
\begin{itemize}
\item better knowledge of the initial state,
\item well defined energy and ability to perform energy scans,
\item much lower 
backgrounds than LHC and therefore the ability to detect signals 
which have low cross sections (e.g. sleptons) or prohibitive backgrounds at
LHC ({\it e.g.}, Higgs bosons decaying hadronically into light quarks),  
\item better measurement of angular distributions and therefore particle 
helicities,
\item polarized beams which allow measurement of quantum numbers, and the 
reduction of major backgrounds (e.g., $WW$).
\end{itemize}

The LC is uniquely capable of measuring the quantum numbers of new particles. 
In this way, the LC can determine the nature and underlying origin of new 
phenomena discovered at the LHC and also provides a unique discovery window 
on its own.  The search reach for new physics via virtual effects at the LC 
exceeds that of the LHC in many scenarios. The LC is both sensitive to new 
physics that LHC cannot observe (or cannot observe well), and can aid LHC in 
distinguishing multiple interpretations of TeV-scale phenomena. These 
capabilities have been detailed in the answer to question 30b. To further 
illustrate this point, we expand on several of the items presented in the 
answer to 30b.

(1) Electroweak Symmetry Breaking:  The LC will precisely measure, at the 
percent level, the properties of the Higgs boson in a model independent way 
and thus experimentally verify the scalar dynamics responsible for electroweak 
symmetry breaking; this is not possible at the LHC for a light Higgs boson. 
For example, for a 120 GeV Higgs boson, the $b\bar b$ (tau, charm, gluon) 
branching fraction can be determined at the level of 
1\% (5\%, 10\%, 10\%) at a 
500 GeV machine. At a 1 TeV LC, the top-quark Yukawa coupling and Higgs 
self-coupling can be measured with an accuracy better than 10\%.  

For strong electroweak symmetry breaking, detailed measurements of cross 
sections and angular distributions at a LC will be essential for identifying 
the new states and disentangling the underlying physics. The 500 GeV LC can 
establish the existence of a new state with a significance better than 5 sigma 
for values of the model parameters in accordance with current constraints; the 
significance increases by more than a factor of two at a  1 TeV machine. In 
addition, the LC can separate the different isospin production channels, such 
as those responsible for the reactions $WW \to WW$ and $ZZ \to WW$,
which is not accessible at the LHC.

(2) Hierarchy Problem:   The prospects that the color neutral part of the  
supersymmetric spectrum (sleptons, charginos, neutralinos) is accessible at  
$500-1000$ GeV are very good.  
The LC can make precise (100 MeV or better) mass  
measurements, as well as coupling, spin, and mixing parameter determinations 
of the supersymmetric partners.  In particular, the accurate mass 
determination of the lightest  supersymmetric particle will sharpen all the 
mass determinations and understanding of superpartner decay chains at the LHC.
These measurements can uniquely confirm the symmetries predicted by 
supersymmetry.  A general exploration of the SUSY breaking mechanism and 
extrapolation to the GUT scale is only possible by combining the data from 
the LC and LHC (see LHC/LC report). 

In the case of extra spatial dimensions, the polarized beams and accurate 
measurement of angular distributions at a LC allow for the simultaneous 
determination of the size, geometry, and number of the additional dimensions 
in a model independent fashion.  This is achievable at a 500 GeV (1 TeV) LC 
for the direct production of gravitons in models with
extra dimensions for a fundamental scale of $3-5$ ($6-10$) TeV. 
In addition, the LC capability for 
the identification of spin-2 exchange in all scenarios can demonstrate the 
connection to gravity.

In any scenario addressing the hierarchy, precision measurements of Standard 
Model processes at the LC with polarized beams are sensitive to virtual 
effects at energy scales significantly beyond 1 TeV and will be crucial to 
determine the nature of the new physics.  Important information will come 
from a 500 GeV LC; running at higher energies will, of course, improve the 
precision and sensitivity to new physics.  For example, limits obtained from 
difermion production on compositeness and extra gauge bosons scale with the 
center-of-mass energy; spin-2 exchange scales somewhat less with energy.

(3) Dark Matter:  Good candidates for the dark matter are neutralinos from 
supersymmetry and Kaluza-Klein excitations of the photon from extra dimension 
theories. A $500-1000$ GeV machine is expected to cover the cosmologically 
favored parameter region within supersymmetry. This affords an intriguing 
opportunity to compare particle accelerator measurements to those from 
astrophysics experiments. In fact, the LC is unique in its capability to 
provide a measurement of the supersymmetric dark matter relic density to a 
precision of 3\% which matches the 
level expected from future astrophysical observations of 2\% (such as PLANCK). 
In addition, the LC is necessary to identify the superpartners which can 
complicate the SUSY dark matter scheme. For instance, if the LSP has a 
slightly heavier partner with the same quantum numbers and a larger 
annihilation cross section, the effective LSP annihilation is significantly 
altered. Simply knowing the LSP mass and self annihilation cross section is 
not enough to place this as the dark matter as precise measurements of the 
full spectrum are needed. If the 
dark matter consists of KK excitations, the LC with its superb energy 
resolution will be essential in identifying the narrow states, and their spin 
can be determined from angular distributions.

4) Precision measurements:  this has been described in the answer to 30b, but 
indeed this is also very important for this question.  The LC capability for 
precision measurements of Standard Model 
processes provides a unique window on new 
physics. The 500 GeV LC has a sensitivity to many processes of new physics 
with a mass reach well beyond that of the LHC in many cases. 

In conclusion, the clean experimental environment of the LC and the unique 
ability to select helicity channels and measure quantum numbers open new 
avenues to discover, identify, and reveal the underlying structure of new 
physics. As for the energy question, we have shown 
that the baseline LC operating at $200-500$ GeV (or as low as 90 GeV as an 
option) is an essential tool for understanding the physics of the TeV scale, 
independent of LHC. Upgrading to 1 TeV opens new potentials for discovery.

\vspace{1.0cm}

\noindent{Acknowledgements}
The work of JLH was supported by the U.S. Department of Energy under contract 
number DE-AC02-76SF00515.  The work of GW was supported by the European
Community's Human Potential Programme under contract HPRN-CT-2000-00149
`Physics at Colliders.'

\end{document}